\documentclass[a4paper]{jpconf}
\usepackage{graphicx}
\usepackage{subfigure}
\usepackage{amsmath}
\usepackage{amssymb}
\usepackage{bm}
\newcommand{\qmoy}[1]{\bigl\langle #1 \bigr\rangle}        
\newcommand{\brkt}[3]{\langle #1| #2 |#3\rangle}                

\newcommand{\hbw}{\hbar \omega}

\newcommand{\s}{1s_{1/2}}
\newcommand{\vlk}{V_{\text{low-}k}}
\newcommand{\rc}{\rho_\pi^{\mathrm{c}}}
\newcommand{\rn}{\rho_\pi^{\mathrm{n}}}
\newcommand{\rmic}{\rho_\pi^{\mathrm{m}}}

\begin{document}
\title{Radii in the $sd$ shell and the  $\s$ ``halo'' orbit: A game changer}
\author{J. Bonnard$^{1,2}$ and A.~P.~Zuker$^{3,4}$}
\address{${}^1$ Institut de Physique Nucl\'eaire, CNRS-IN2P3, Universit\'e Paris-Sud, Universit\'e Paris-Saclay, 91406 Orsay Cedex, France}
\address{${}^2$ Istituto Nazionale di Fisica Nucleare, Sezione di Padova, 35131 Padova, Italy}
\address{${}^3$ Dipartimento di Fisica e Astronomia, Universit\`a degli Studi di Padova, I-35131 Padova, Italy}
\address{${}^4$ Universit\'e de Strasbourg, IPHC, CNRS, UMR7178, 23 rue du Loess 67037 Strasbourg, France}
%
%
\begin{abstract}
A new microscopic parametrisation of nuclear radii as a functional of single-particle occupation numbers is presented.  Its form is inspired
by the Duflo-Zuker phenomenological fit which contains a ``correlation'' term that recently made it possible to understand the
isotope shifts of several species as due to unexpectedly large $\s$ and $1p$ orbits [Bonnard J, Lenzi S M and Zuker A P 2016 \textit{Phys.Rev. Lett.}  \textbf{116} 212501].  
It will be shown that the calculated radii for $sd$-shell nuclei reproduce the
experimental data better than the most accurate existing fits. These results reveal a very peculiar behaviour of the $\s$ orbit: It is huge
(about 1.6 fm bigger than its $d$ counterparts of about 3.5 fm) up to $N, Z=14$, then drops abruptly but remains some 0.6 fm larger than the $d$ orbits. 
An intriguing mechanism bound to challenge our understanding of shell formation.
\end{abstract}

\section{Introduction}
Proton radii of nuclei in the $sd$ shell depart appreciably from the asymptotic $\rho_0A^{1/3}$ law.  The discrepancy exhibits systematic
trends fairly well described by a single phenomenological term in the Duflo-Zuker (DZ) formulation \cite{DZIII}, which also happened to
explain the sudden increase in slope in the isotope shifts of several chains at neutron number $N=28$ \cite{haloskins,Mn}. It was shown that
this term is associated to the filling of abnormally large $\s$ and $1p$ orbits in the $sd$ and $pf$ shells respectively. 

Further to explore the question of ``halo'' orbits, here we propose a microscopic calculation of $sd$-shell nuclei radii.  As the (square)
radius is basically a one body operator, it can be represented as a functional of single-particle occupancies which will be determined by means of
the interacting shell model \cite{rmp}. The only adjustable parameter models a correction to the mean-field radius of the $\s$ orbit
whose evolution will be optimized by demanding agreement of the computed radii with experimental values. 

In next section we briefly review the DZ fitting formula giving particular attention to the term that led to the detection of ``halo''
orbits in Ref. \cite{haloskins}. Section~\ref{calc} starts by introducing the necessary definitions and clarifying some delicate
conceptual points (Subsection~\ref{defcon}) Then comes subsection~\ref{mci} describing a new shell-model interactions used in
subsection~\ref{microsec} where the microscopic parametrisation is discussed and the fundamental results obtained. Finally,
Section~\ref{concsec} is devoted to explain why these results amount to a ``game changer''.
\section{The Duflo-Zuker phenomenological formula} \label{DZsec}
As self-bound systems, nuclei have volumes that go as the number of particles $A$.  Therefore their radii go as $A^{1/3}$.  As both
neutrons and protons are present, an isospin dependence is also expected.  Duflo and Zuker (DZ) \cite{DZIII} proposed the followings
($t_z=N-Z$, $\rn$ and $\rc$ stand for naive and correlated radii respectively):
\begin{gather}
\rc= \rn + \lambda \left( \dfrac{z(D_\pi-z)}{D_\pi^2}\times \dfrac{n(D_\nu-n)}{D_\nu^2}  \right )A^{-1/3}, \label{corrfit} \\
\rn = A^{1/3} \left( \rho_0-\frac{\zeta}{2}\frac{t_z}{A^{4/3}}  - \frac{\upsilon}{2}\Bigl(\frac{t_z}{A}\Bigr)^2\right)e^{g/A}, \label{radii}
\end{gather}
where $n$ ($z$) is the number of active neutrons (protons) between the extruder-intruder (EI) magic numbers \cite{rmp} $N,\, Z= 6, 14, 28, 50\ldots$,
while $D_x=8, 14, 22\ldots $are the corresponding degeneracies.  In Eq.\eqref{radii}, the exponential term accounts for the larger size of light nuclei, $\upsilon$ measures the overall dilation or contraction as a function of $t_z^2$, and $\zeta$ is responsible for the difference in radii between the fluids, i.e. the neutron skin.  Indeed, assuming isospin conservation implies that the neutron radius of a nucleus equals the proton radius of its mirror: $\rho^{\mathrm{n,\,c}}_\nu (A,t_z)=\rho^{\mathrm{n,\,c}}_\pi(A,-t_z)$, which yield for the neutron skin thickness $\Delta r_{\nu\pi}=\rho^{\mathrm{n,\,c}}_\nu-\rho^{\mathrm{n,\,c}}_\pi=\frac{\zeta}{t}{A} e^{g/A}$. The quality of the results depends little on $\zeta$ in the range $0.0<\zeta<1.2$ fm.  The chosen value $\zeta=0.8$ fm represents the DZ estimate for $\Delta r_{\nu\pi}$ that agrees very
well with \textit{ab initio} and experimental values \cite{haloskins}.

The DZ fits reproduce much better the isotope shifts $\delta\qmoy{r^2_\pi}$---relative to a reference nucleus---than the absolute value of the latter which demands reexamination, carried out only in the original DZ paper. We shall consider two data sets:
\begin{itemize}
\item[$\bullet$] The compilation used in DZ \cite{DZIII}, labelled D. It is based on data used in Duflo's original work~\cite{duflo} and (mostly) on the tables of Najdakov and coworkers~\cite{nadjakov}. To ensure consistency with the isotope shifts 15 absolute values were corrected. 
\item[$\bullet$] The recent Angeli-Marinova (AM) compilation \cite{AM}.
\end{itemize}
\begin{table}[b!]
\caption{\label{tab:fits}Results of fits for D and AM: poor for $\rn$ from Eq.~\eqref{radii} (rows 1,5-3,7), improve radically for $\rc$ from Eq.\eqref{corrfit} (rows 2,6-4,8). There are $\nu$ data points for $Z\leq Z_M$ and $N\geq Z$. Row 9 keeps only those AM values for which isotope shift measures are available. $\zeta=0.8$ fm is fixed, $\rho_0$, $\lambda$, $\upsilon$, and root-mean-square deviations (rmsd) in fm.}
\begin{tabular*}{\linewidth}{@{\extracolsep{\fill}}ccccccccc}
\hline
\hline
I&$\rho_0$& $g$& $\lambda$& $\upsilon$& Data&$\nu$& $Z_M$&rmsd\\ 
\hline
1 &0.946&  1.422&  - &  0.550& D & 636&  96& 0.0299\\
2 &0.942&  0.948&  6.857& 0.526&  D & 636&  96& 0.0132\\ 
3 &0.950&  1.232&  - &  0.312& D &  88&  30& 0.0415\\
4 &0.944&  0.985&  5.562& 0.368&  D &  88&  30& 0.0176\\
\hline
5&0.942&  1.513&  - &  0.312& AM& 876&  96& 0.0365\\
6&0.940&  0.879&  7.719& 0.297&  AM& 876&  96& 0.0203\\
7&0.947&  1.370&  - &  0.295& AM& 107&  30& 0.0419\\
8&0.940&  1.140&  5.208& 0.334&  AM& 107&  30& 0.0246\\
\hline
9&0.930&  1.659&  4.499& 0.379&  AMr2&80&  30& 0.0176\\
\hline
\hline
\end{tabular*}
\end{table}
The numbers speak by themselves: the $\lambda$ contribution (the Duflo term from now on) makes an enormous difference. Its beneficial effects
extend to all regions. In particular in Ref.~\cite{haloskins} it explains the---hitherto puzzling---increase of slope in the Ca and K
isotope shifts \cite{expCaK1,expCaK2}. Row 9 in the table is of particular interest: leaving out absolute radii unrelated to isotope
shifts reduces very significantly the rmsd in row 8. The necessary reexamination of the AM set is under way~\cite{onrad}.

\section{Microscopic approach}\label{calc}
\subsection{Definitions and conceptual points}\label{defcon}
\begin{figure}[b]
\begin{center}
\includegraphics[angle=-90,width=0.5\textwidth]{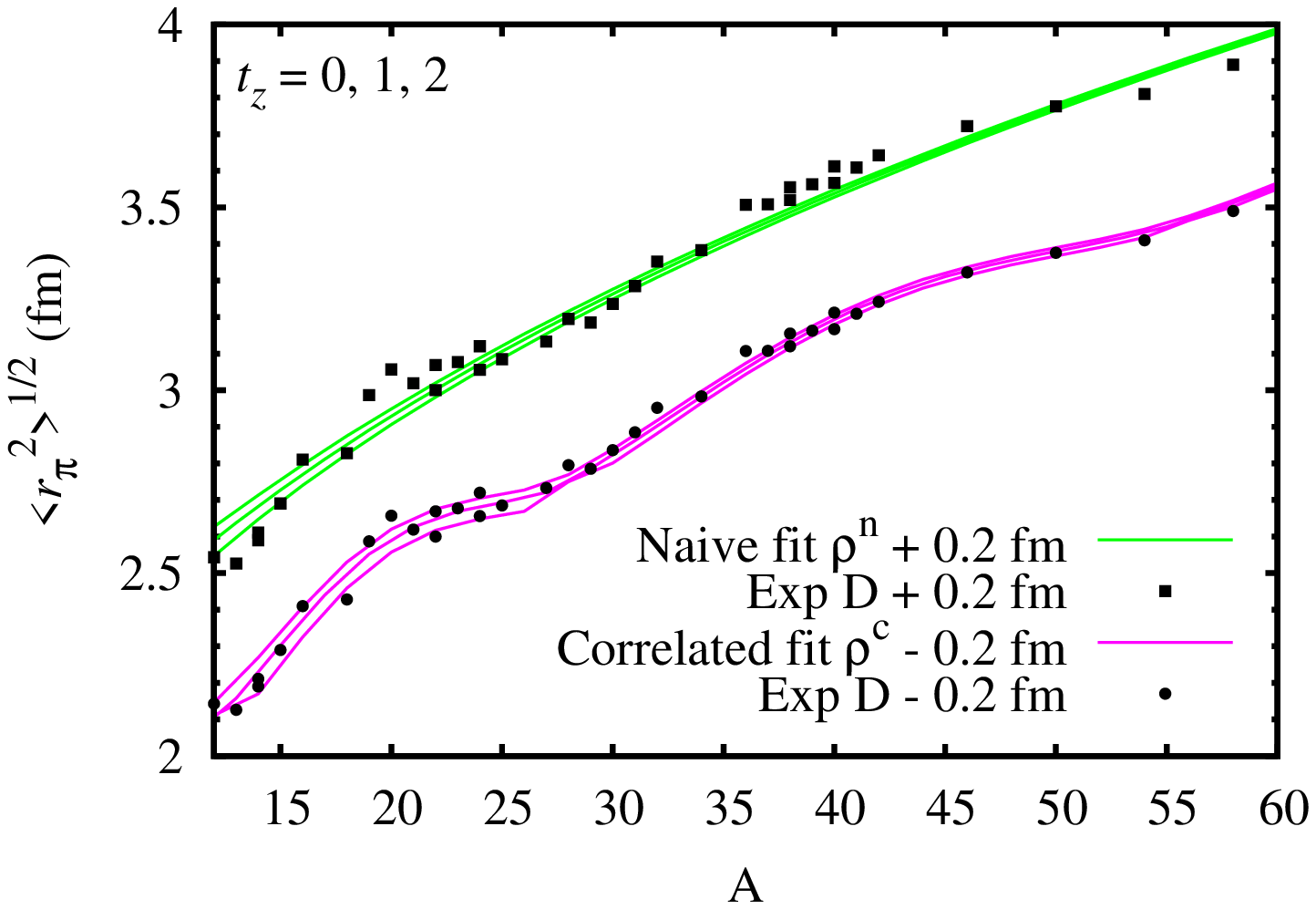}%
\includegraphics[angle=-90,width=0.5\textwidth]{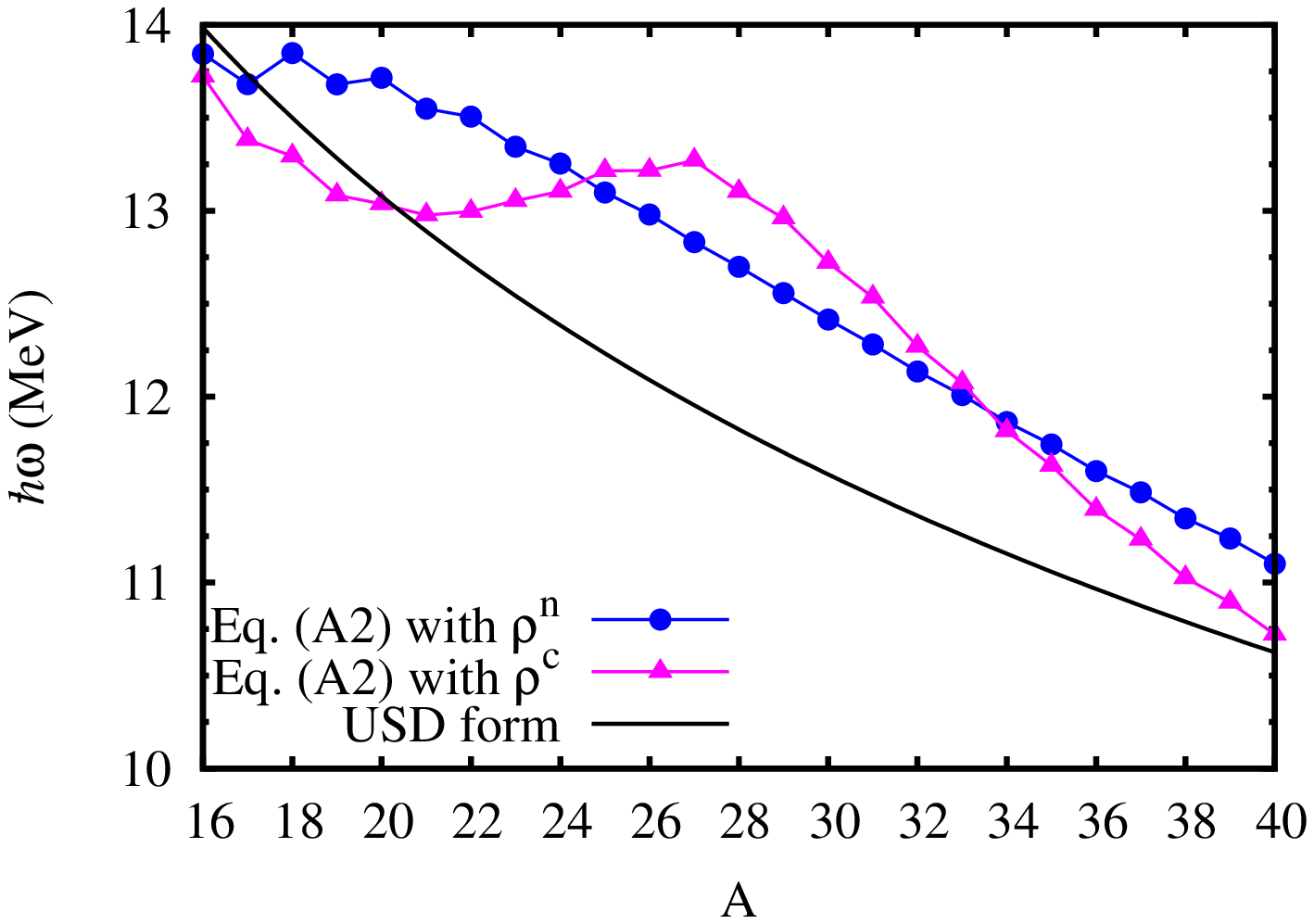}%
\caption{(color online) Left panel: The influence of the Duflo $\lambda$ term. Fits of the D data, for nuclei with $t_z=0,1,2$ and $Z\le 30$, using $\rn$ and $\rc$, labeled naive and correlated respectively. Right panel: Evolution of $\hbw$ for naive, correlated and USD assumptions.  Only $t_z=0$ and 1 cases are included, merged in a single curve which leads to the weak staggering observed.\label{fig:rhw}}
\end{center}
\end{figure}

Let us consider two square radius operators defined through their expectation values in a stationary state as
\begin{subequations}\label{op}
\begin{align}
&\qmoy{r_{\mathrm{ho}}^2 }=\frac{41.47}{\hbw} \sum_i m_i(p_i+3/2)/A \quad \quad\quad(\sim (\rn)^2)\\
&\qmoy{r_\pi^2}=\frac{41.47}{\hbw}\sum_i m_i(p_i+3/2+\delta_i)/A    \quad\:\:(\sim (\rc)^2), \label{r2}
\end{align}
\end{subequations}
where $m_i$ and $p_i$ represent the total occupancy and principal quantum number of the orbit $i=(p_il_ij_i)$, and $41.47(p_i+3/2)/\hbw$ fm$^2$ is the harmonic oscillator (ho) value of the square radius for any single-particle wave function belonging to the major shell.
Such estimates of radii provide a direct reading of the naive and correlated ($\rn$ and $\rc$) patterns shown in the left panel of Figure~\ref{fig:rhw} that illustrates the improvement brought about by Duflo's {$\lambda$} term. Its effects are identified with corrections $\delta_i$ to the radii of harmonic orbitals. Note that under this guise, i.e. without any additional assumption regarding the $\delta_i$, $r_{\pi}^2$ is a fully general expression for the proton square radius operator.

Eqs. \eqref{op} induce two possible expressions for the oscillator frequency
\begin{subequations}\label{hbw}
\begin{align}
\label{stdhw} &\hbw_x=\frac{41.47}{\qmoy{r_x^2}}\sum_i m_i(p_i+3/2)/A \\  
\label{nstdhw} &\hbw=\frac{41.47}{\qmoy{r_{\pi}^2}}\sum_i m_i(p_i+3/2+ \delta_i)/A,
\end{align}
\end{subequations}
with $x$ a generic index, e.g. $x= $ ho.
$\hbw$ enters shell-model calculations through the two-body matrix elements, taken to scale as $V(\hbw)=V(\hbw_r)\hbw/\hbw_r$ where $\hbw_r$ is an arbitrary reference value. Its estimate will depend on the choice of the $\qmoy{r^2_x}$ denominator. For $x= $ ho, an asymptotic estimate of the an asymptotic estimate of the sum \eqref{stdhw} with the very schematic estimate $\qmoy{r^2}\propto A^{1/3}$ leads to $\hbw=35.6A^{1/3}/\qmoy{r^2}\approx 40A^{-1/3}$, the classic result \cite[Eq.(2.157)]{BM} on which shell-model calculations traditionally rely to ensure correct saturation properties.
A more precise alternative is a nucleus by nucleus evaluation of the sum while resorting to the naive fit Eq.~\eqref{radii}, $\qmoy{r^2_x}\equiv (\rn)^2$. A more audacious one relies on Eq.~\eqref{corrfit}. It amounts to assume that the effect of the $\delta_i$ modifications (see Eq. \eqref{nstdhw}) is ``transferred'' to $\hbw_x$ by spreading it among oscillator, which is achieved by replacing in Eq. \eqref{stdhw} the natural $\qmoy{r^2_\mathrm{ho}}$ denominator by the (closer to) exact one $\qmoy{r^2_{\pi}}$ approximated \textit{via} the correlated radii \eqref{corrfit}. 
The two variants are plotted in the right panel of Fig.~\ref{fig:rhw}, where the USD form $\hbw(A)=\hbw(A=18) (18/A)^{0.3}$ is also shown~\cite{USD1}. The calculations will be done with both the USDa interaction and the correlated form.
Both the USD-like and correlated propagation will be considered in the following.

The use of the notation $\qmoy{r^2_{\pi}}$ stresses the obvious: We are dealing exclusively with proton radii. What may not be so obvious is that
Eq.~\eqref{r2} is written in terms of $m_i=n_i+z_i$ and no reference whatsoever is made to $z_i \text{ or } n_i$, the number of protons or
neutrons, while Eq.~\eqref{radii} does contain a term in $t_z=N-Z$. However (as already shown in~\cite{DZIII}) for any value of
$\zeta$ in Eq.~\eqref{radii} one could find a value of $\upsilon$ associated to $t_z^2$ that would produce the same rmsd. Furthermore,
the all important Duflo term is an isoscalar. Finally the neutron an proton radii are necessarily very close: Otherwise it would be
impossible to explain why the structure of the isotope shifts are dictated by the {\em neutron} filling patterns, while what is measured
are radii for {\em protons} whose number remain fixed~\cite{haloskins}. Hence we {\em must} use an isospin representation. But this means that the expression for $\qmoy{r^2_{\pi}}$ in Eq.~\eqref{r2} must be supplemented by an isovector term in $(t_z)_i=n_i-z_i$, which---at this point---cannot be
determined, and is therefore ignored: A limitation of this study, unlikely to alter the central result of Section~\ref{microsec} as such term contribute at next order only.

\subsection{The monopole corrected interaction}\label{mci}
This section is included for the sake of completeness. Indeed, it needs revisiting as our view of the corrective action demanded by realistic
interactions will be affected by the findings in next section.

The monopole corrected interaction MCI is derived from the N3LO potential~\cite{N3LOa} at $\hbw_0=14$ MeV, and a $\vlk$ treatment~\cite{vlk} with the cut-off $\lambda=2\text{fm}^{-1}$ in momentum space, as part of the no-core project outlined in Ref.~\cite{haloskins}. Renormalization amounts to an overall 1.1 multiplicative factor and a 30\% boost of the quadrupole force~\cite{mdz} plus the monopole corrections defined below.  The whole is scaled as mentioned previously with $\hbw$ from the correlated form in the right panel of Fig.~\ref{fig:rhw}.

\begin{figure}[t]
\begin{center}
\begin{minipage}[!b]{0.42\textwidth}
\includegraphics[width=\textwidth]{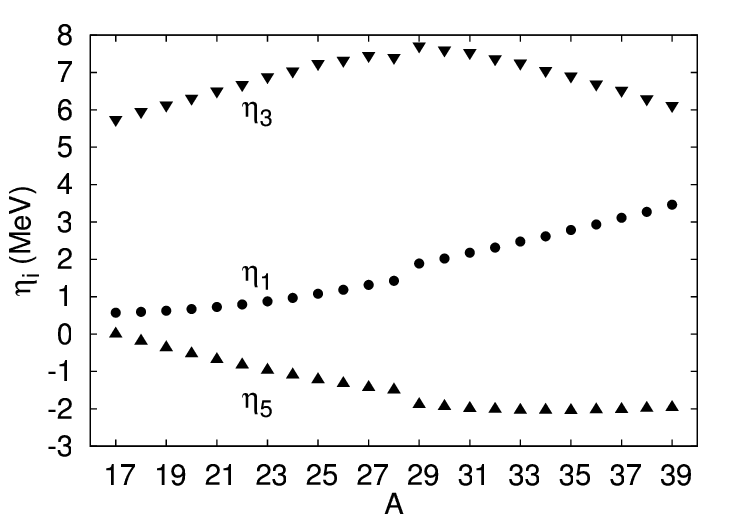}%
\end{minipage}\hspace{0.0\textwidth}
\begin{minipage}[!b]{0.48\textwidth}\vspace{2.cm}\caption{\label{fig:eta} Propagation of the MCI single-particle energies $\eta_i$ from the first line of Eq.~\eqref{Monop}.}
\end{minipage}
\end{center}
\vspace{-0.95cm}
\end{figure}

The monopole corrections are given in terms of operators in the ``invariant representation''~\cite{acz}, in which the one- and two-body number operators $m_t$ and $m_tm_u$ are separated into a term that contains only the total number operator $m$, single-particle terms $\Gamma^{(1)}_t$, and two-body terms $\Gamma^{(1)}_{tu}$:
\begin{equation}
m_t\equiv m+\Gamma^{(1)}_{t}, \quad
\frac{m_t(m_u-\delta_{tu})}{1+\delta_{tu}}\equiv
\frac{1}{2}m^{(2)}+(m-1)\Gamma^{(1)}_{t}+\Gamma^{(2)}_{tu},
\end{equation}
where $x^{(k)}=x(x-1)\dots(x-k+1)$.  For our purpose it is convenient to associate each orbit $t$ with its complement $ct$ containing all orbits in the space except $t$.  The degeneracy of $t$ is $D_t=2(2j_t+1)$ and that of its complement $D_{ct}=D-D_t$, $D$ being the total degeneracy of the valence space.  In the $sd$ shell, orbits $d_{5/2},\, s_{1/2}$ and $d_{3/2}$ will be called $5,\, 1,\, 3$ respectively. The complement of $5$, say, is $c5=1+3$.  With this convention:
\begin{equation}
\Gamma_{t}^{(1)}=\frac{m_tD_{ct}-m_{ct}D_{t}}{D},\quad
\Gamma_{t}^{(2)}=\left(\frac{m_t^{(2)}}{
  D_t^{(2)}}+\frac{m_{ct}^{(2)}}{D_{ct}^{(2)}} -\frac{2m_tm_{ct}}{
  D_tD_{ct}}\right) \frac{D_t^{(2)}D_{ct}^{(2)}}{(D_t+D_{ct})^{(2)}}.
\end{equation}
The monopole corrections finally take the form
\begin{align} \label{Monop}
  V_\text{MC} =e_5\Gamma_5^{(1)}(m-1)+0.06\Gamma_1^{(1)}(m-1) &+
  2.0\Gamma_3^{(1)}(m-1)(\overline{m}-1)/(D/2-1)^2 \\ &+
  (1.0\Gamma_5^{(2)}+0.5\Gamma_3^{(2)}) \overline{m}/D, \label{Mono3}
\end{align}
where $e_5=-0.05$ if $ A\le 28$, $-0.11$ if $A>28$. All coefficients in MeV.  The sum in the first line defines propagated single-particle operators $\eta_t$ such that $\sum_t \eta_t D_t=0$ (see Fig.~\ref{fig:eta}). Note that the terms containing $\overline{m}-1$ and $\overline{m}$ amount to a three body contribution. Under particle-hole transformations $m_t\to \overline{m_t}=D_t-m_t$, $\Gamma_{t}^{(1)}$ changes sign and $\Gamma_{t}^{(2)}$ is invariant.

The description of spectra is quite satisfactory for the yrast states, not as good as USDa for the rest, with the exception of ${}^{24}$Mg and ${}^{28}$Si where MCI does better. Figure~\ref{fig:eta} indicates that the fit detects a---premonitory---abrupt change at $A=28$.
%
%
\subsection{Microscopic parametrisation of nuclear radii} \label{microsec}
\begin{figure}[!b]
\begin{center}
\begin{minipage}{0.42\textwidth}
\includegraphics[angle=-90,width=\textwidth]{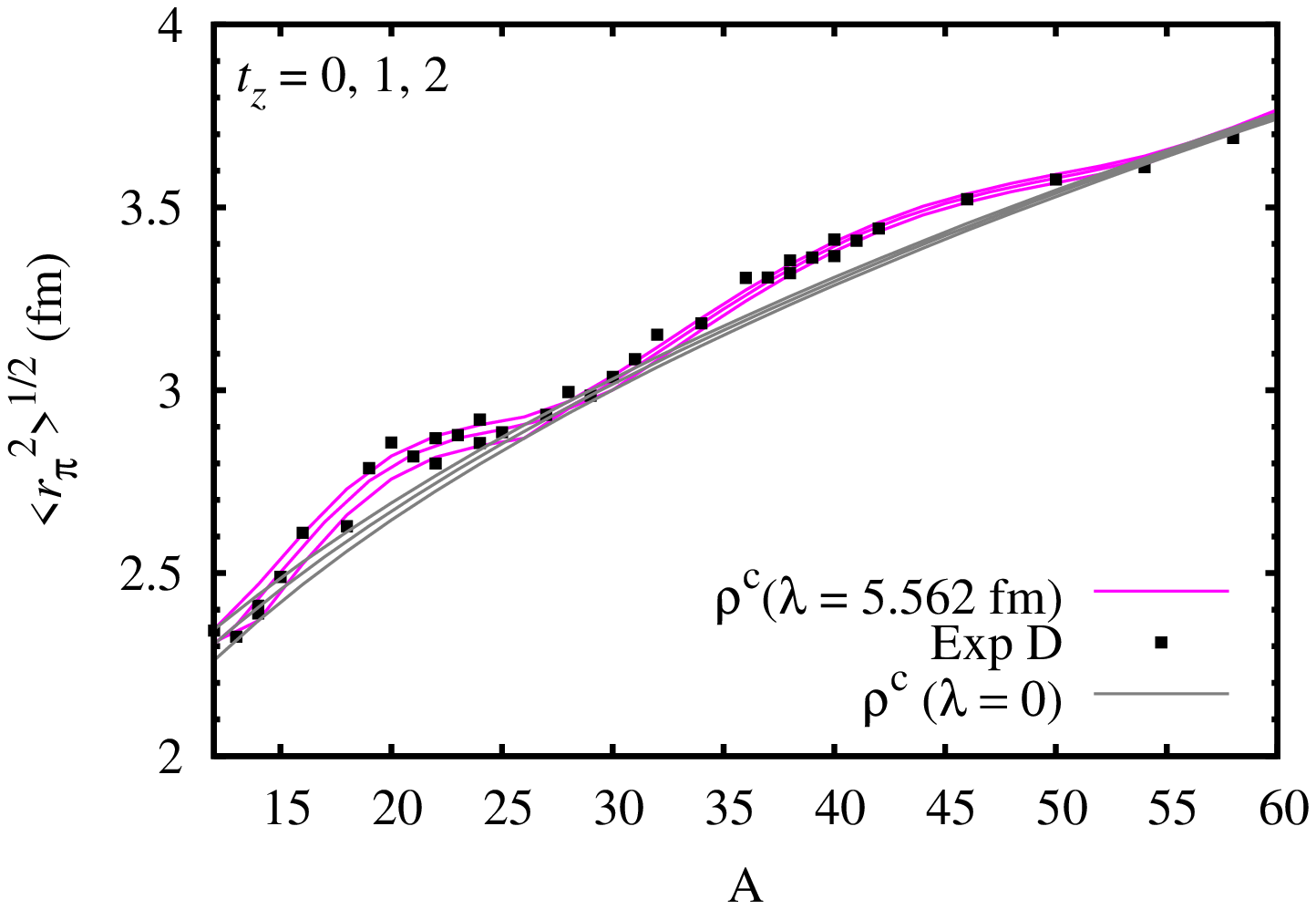}%
\end{minipage}\hspace{0.0\textwidth}
\begin{minipage}{0.48\textwidth}\vspace{1.3cm}\caption{\label{fig:DDZtan}(color online) As Fig.~\ref{fig:rhw} (left panel) with the naive fit
 ($\lambda=0$) replaced by the result of using the parameters of the correlated fit setting $\lambda=0$, which ensures that the corrective term is definite positive.}
\end{minipage}
\end{center}
\vspace{-0.95cm}
\end{figure}

We know that the origin of the Duflo term, Eq. \eqref{corrfit} must be found in the ``halo'' orbits. Concentrating on the $sd$ shell, the $\s$ orbit is taken to be solely responsible for the fluctuations. Accordingly, setting $\delta_i=0$ for $p_i=0d_{5/2,3/2}$ and $\delta_i=\delta$ for $i=\s$, and following Eq. \eqref{r2}, the proposed functional reads
\begin{equation} \label{r2pi}
  (\rmic)^2=\frac{41.47}{\hbw_0}\sum_i m_i(p_i+3/2+\delta_i)/A= \frac{41.47}{\hbw_0}\sum_i
  m_i(p_i+3/2)/A+\frac{41.47}{\hbw_0}m_{s_{1/2}}\delta_i/A
\end{equation}
To determine $\hbw_0$ we refer to Fig.~\ref{fig:DDZtan} where the naive part is not obtained by fitting Eq.~\eqref{radii} but Eq. \eqref{corrfit} and keeping only the $\lambda=0$ part, i.e. with $\qmoy{r_x^2}=[\rc(\lambda=0)]^2$ in Eq. \eqref{stdhw}. Therefore, \eqref{r2pi} reduces to
\begin{equation}\label{r2d}
(\rmic)^2=\rc(\lambda=0) +\frac{41.47}{\hbw_0}\frac{m_{s_{1/2}}\delta}{A}.
\end{equation}
At this stage, the evolution of $\delta$ with $A$ and $t_z$ may be extracted from a nucleus-by-nucleus adjustment.  Here, instead, we simply adopt a step function discontinuous at the EI closure:
\begin{gather}
\delta=
\left\lbrace
\begin{array}{lccl}
\delta_<  & \text{if} & N \text{ and } Z & <14 \\
\delta_>  & \text{if} & N \text{  or } Z & \geq 14.
\end{array}\right.
\end{gather}
Finally, the occupancies are determined by diagonalization of the USDa and MCI Hamiltonians.

\begin{figure*}[t]
\begin{center}
\subfigure{\hspace{-1cm}
\includegraphics[angle=-90,width=0.53\textwidth]{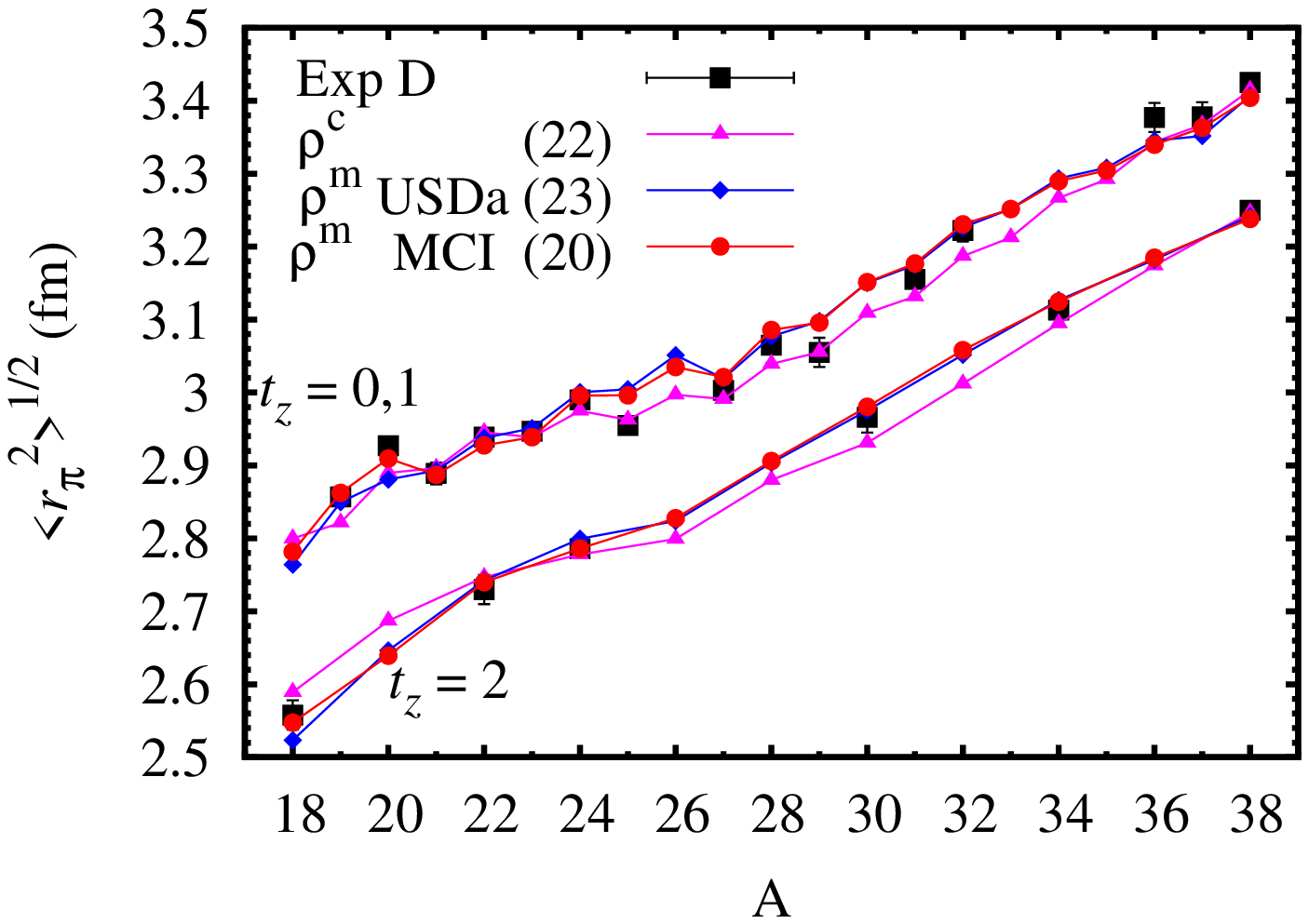}\hspace{-1.9cm}
\includegraphics[angle=-90,width=0.53\textwidth]{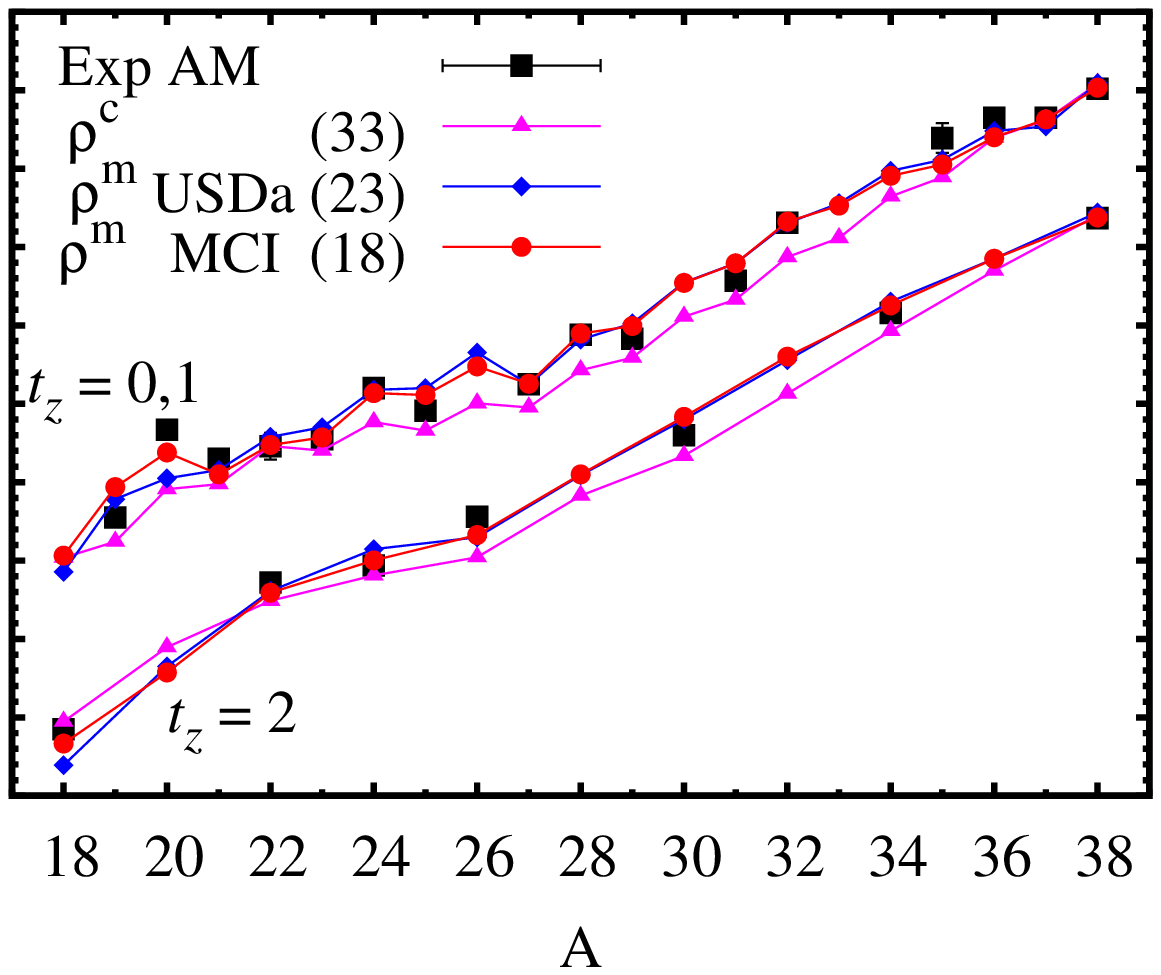}}
\vspace{-0.7cm}
\end{center}
\caption{\label{fig:resDAM}(color online) Solution of Eq.~(\ref{r2pi}): Behaviour of $\rmic$ for USDa and MCI calculations compared with data sets D (left), AM (right), and associated correlated fits $\rc$. The $t_z=0$ and 1 cases are conflated in a single curve. These and $t=2$ values are shifted by $\pm0.07$ fm, respectively, for clarity. In parenthesis the rmsd ($10^{-3}$ fm) for the different cases (see Table~\ref{tab:delta}).}
\vspace{-0.5cm}
\end{figure*}

\begin{figure}[b]
\begin{center}
\begin{minipage}[!b]{0.47\textwidth}
\includegraphics[angle=-90,width=\textwidth]{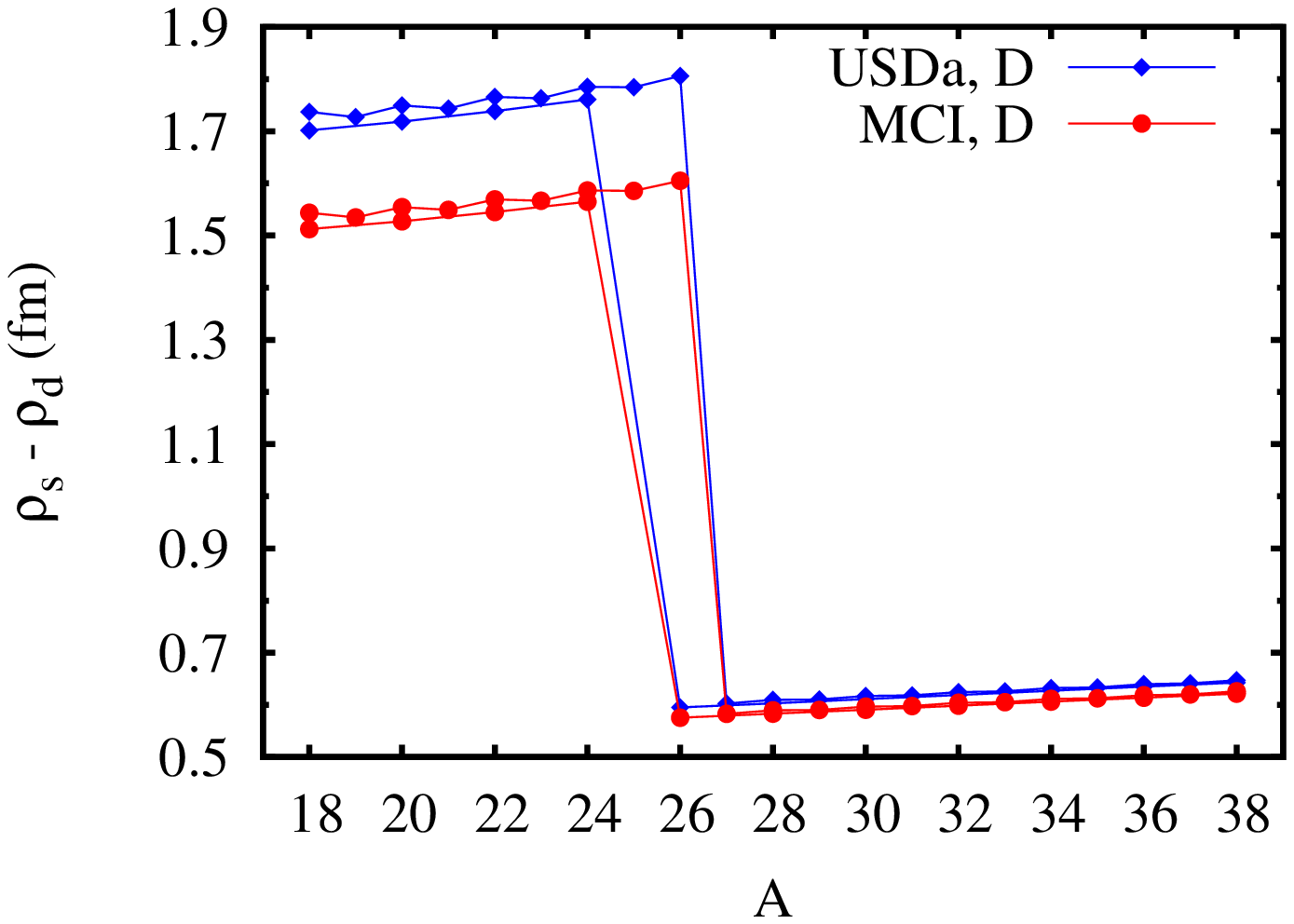}%
\end{minipage}\hspace{0.0\textwidth}
\begin{minipage}[!b]{0.45\textwidth}\vspace{-0.55cm}\caption{\label{fig:rsrd}(color online) Differences $\rho_s-\rho_d$, where $\rho^2_i=\brkt{im_{j_i}}{r^2}{im_{j_i}}=\frac{41.47}{\hbw_0}(p_i+3/2+\delta_i)$, between the root-mean-square radii of the $\s$ and $0d$ orbits from Eq.~\eqref{r2d}. The mean values (in fm) associated to the upper and lower $\delta$ in Table~\ref{tab:delta} are (1.56, 0.6) for MCI D, (1.75, 0.62) for USDa D (shown); and (1.76,0.6) for MCI AM, (1.93,0.69) for USD AM (not shown).}
\end{minipage}
\end{center}
\vspace{-0.9cm}
\end{figure}

A search for the optimal parameters minimizing the rmsd of radii yields the values in Table~\ref{tab:delta} leading to Figures~\ref{fig:resDAM}, in which the microscopically calculated radii are compared with the phenomenological ones obtained with the Duflo term. It appears that the microscopic results reproduce the observed radii at least as well as the phenomenological fit for the D set, and much
better in the AM case. The rmsd for MCI is smaller than for USDa, and so are the calculated sizes of the $\s$ orbit deduced from Table~\ref{tab:delta} and shown in Figure~\ref{fig:rsrd}. The quantitative differences between the results from the different interaction are not really relevant when compared to the fundamental
message.
\begin{table}[!t]
\begin{center}
\begin{minipage}[!b]{0.5\textwidth}
\begin{tabular*}{\linewidth}{@{\extracolsep{\fill}}ccccc}
\hline
\hline
Exp. set      &\multicolumn{2}{c}{Duflo}&\multicolumn{2}{c}{Angeli-Marinova} \\
\hline
Interaction   &  USDa   &   MCI     & USDa    &  MCI   \\
$\delta_<$ &  4.90   &  4.25     &  5.50   &  4.80 \\
$\delta_>$ &  1.40   &  1.35     &  1.45   &  1.35 \\
 rmsd (fm) &  0.023  &  0.020    &  0.023  &  0.018 \\
\hline
\hline
\end{tabular*}
\end{minipage}\hspace{0.01\textwidth}
\begin{minipage}[!b]{0.41\textwidth}\vspace{-0.3cm}\caption{\label{tab:delta} Results for the optimal $\delta$'s for D and AM sets of data, and
for the USDa and MCI interactions. The rmsd are calculated for $sd$-shell nuclei with $t_z=0, 1,$ and 2 (21 in D, 23 in AM).}
\end{minipage}
\end{center}
\vspace{-0.8cm}
\end{table}

\section{Conclusion: The game changer}\label{concsec}
The halo orbits are the essential ingredient in halo nuclei \cite{Tanihata-halos}. What we are finding is that they have a pervasive influence throughout. Their description will most likely require non-local potentials~\cite{perey-buck,Rawitscher-nonlocal}.  At present only relativistic mean-field calculations seem capable of reproducing the slope increase in isotope shifts (as in Ref.~\cite{expCaK1} using the DD-ME2 interaction\cite{rhfDD_ME2}). 

The shell-model theory is conducted under the tacit assumption that harmonic oscillator wave functions approximate well enough the single-particle behaviour, which is tenable for conventional Hartree-Fock or Woods-Saxon calculations, but not for halo orbits. This may explain why some USD matrix elements are impossible to reproduce using perturbation theory on realistic ones calculated with oscillator orbits, in particular for the $JT=20$ state~\cite{mdz} or the coefficient of the $\Gamma_5^{(2)}$ term in Eq. \eqref{Mono3}.

Which brings us to the game changer: The fundamental problems of saturation and shell formation may not be due to limitations of the potentials but to limitations of the calculations that fail to produce the halo orbits. To fix ideas: Why in our N3LO matrix elements do we use a cut-off $\lambda=2\text{fm}^{-1}$ and not $\lambda=4\text{fm}^{-1}$? Because then the calculations would become very hard. A problem that is
ignored, but not solved with existing, Brueckner or $\vlk$, approaches.

Therefore, could it not be that the observed magic closures are not associated to self-binding of the orbits of largest $j$ in a major oscillator shell---due to three-body forces---but, instead, to unbinding of the ``halo'' orbits above them? It sounds improbable, but as Sherlock Holmes put it {\em Once you eliminate the impossible, whatever remains, no matter how improbable, must be the truth.}
%
%
%
\section*{References}
%
%

\end{document}